\documentclass[11pt,onecolumn]{IEEEtran}
\usepackage{graphicx}
\usepackage{subfigure}
\usepackage{epsfig}
\usepackage{amssymb}
\usepackage{amsmath}
\usepackage{amsfonts}

\newcommand{\vp}{\mbox{${\bf p}$}}

\newcommand{\vx}{\mbox{${\bf x}$}}

\newcommand{\vs}{\mbox{${\bf s}$}}

\newcommand{\vn}{\mbox{${\bf n}$}}

\newcommand{\mB}{\hbox{{\bf B}}}

\newcommand{\mC}{\hbox{{\bf C}}}

\newcommand{\mH}{\hbox{{\bf H}}}

\newcommand{\ga}{\alpha}
\newcommand{\gb}{\beta}

\newcommand{\gr}{\rho}

\newcommand{\gD}{\Delta}

\def\bm#1{\mbox{\boldmath $#1$}}

\newcommand{\cR}{{\cal R}}

\newcommand{\SNR}{\ensuremath{\hbox{SNR}}}

\newtheorem{theorem}{Theorem}[section]
\newtheorem{lemma}[theorem]{Lemma}

\newtheorem{prop}{Proposition}[section]
\newtheorem{claim}{Claim}[section]

\newtheorem{definition}{Definition}[section]
\newtheorem{question}{Question}[section]
\newtheorem{coro}{Corollary}[section]

\newcommand{\beq}{\begin{equation}}
\newcommand{\eeq}{\end{equation}}
\newcommand{\bea}{\begin{array}}
\newcommand{\ena}{\end{array}}
\newcommand{\bds}{\begin {itemize}}
\newcommand{\eds}{\end {itemize}}
\newcommand{\bdf}{\begin{definition}}
\newcommand{\blm}{\begin{lemma}}
\newcommand{\edf}{\end{definition}}
\newcommand{\elm}{\end{lemma}}
\newcommand{\bthm}{\begin{theorem}}
\newcommand{\ethm}{\end{theorem}}
\newcommand{\bprp}{\begin{prop}}
\newcommand{\eprp}{\end{prop}}
\newcommand{\bcl}{\begin{claim}}
\newcommand{\ecl}{\end{claim}}
\newcommand{\bcr}{\begin{coro}}
\newcommand{\ecr}{\end{coro}}
\newcommand{\bquest}{\begin{question}}
\newcommand{\equest}{\end{question}}

\newcommand{\rarrow}{{\rightarrow}}

\begin{document}
\title{Bargaining Over the Interference Channel}
\author{Amir Leshem and Ephraim Zehavi
\thanks{School of Engineering, Bar-Ilan University, Ramat-Gan, 52900,
Israel.  Contact Author: Amir Leshem, e-mail:
leshema@eng.biu.ac.il. }}
\date{}
\maketitle
\begin{abstract}
In this paper we analyze the interference channel as a conflict situation. 
This viewpoint implies that certain points in the rate region are unreasonable 
to one of the players. Therefore these points cannot be considered achievable 
based on game theoretic considerations. 
We then propose to use Nash bargaining solution as a tool that provides 
preferred points on the boundary of the game theoretic rate region. We provide 
analysis for the 2x2 intereference channel using the FDM achievable rate 
region. We also outline how to generalize our results to other achievable 
rate regions for the interference channel as well as the multiple access 
channel. 

\end{abstract}
\section{Introduction}
Computing the capacity region of the interference channel is an open problem in information 
theory \cite{cover}. 
A good overview of the results until 1985 is given by van der Meulen  \cite{Meulen94}
and the  references therein. The capacity region of general interference case 
is not known yet. However, in the last forty five years of research some 
progress has been made. Ahslswede \cite{ahlswede73}, derived a general 
formula for the capacity region of a discrete memoryless  Interference 
Channel (IC) using a limiting expression which is computationally infeasible. 
Cheng, and  Verdu \cite{cheng93} proved that the 
limiting expression cannot be written in general by a single-letter formula 
and  the restriction to Gaussian inputs provides only an inner bound to the 
capacity region of the IC. The best known achievable region for the general 
interference channel is due to Han and Kobayashi  \cite{han81}. However the 
computation of the Han and Kobayashi formula for a general discrete memoryless 
channel is in general too complex. Sason \cite{sason2004} describes certain improvement over the Han Kobayashi rate region in certain cases.
In this paper we focus on the 2x2 memoryless 
Gaussian interference channel.  A 2x2 Gaussian interference channel 
in standard form (after suitable normalization) is given by:
\beq
\label{standard_IC}
\vx=\mH \vs +\vn
\eeq
where 
\[
\mH=\left[
\bea{cc}
1 & \ga \\
\gb & 1
\ena
\right]
\]
$\vs=[s_1,s_2]^T$, and $\vx=[x_1,x_2]^T$ are sampled values of the input and 
output signals, respectively. The noise vector $\vn$ represents the 
 additive Gaussian noises with zero mean and unit variance. The powers of the 
input signals are constrained to be less than $P_1,P_2$ respectively. The off-diagonal elements of $\mH$, $\ga,\gb$ represent the degree of interference 
present. The capacity region of the Gaussain interference channel with very 
strong interference (i.e., $\ga \ge 1+P_1$, $\gb \ge 1+P_2$ ) 
was found by Carleial
given by
\beq
\label{VSI_RR}
R_i \le \log_2(1+P_i), \ \ i=1,2.
\eeq
This surprising result shows that very strong interference dose not reduces the capacity. 
A Gaussian interference channel is said to 
have strong interference if  $\min\{\ga,\gb\}>1$. Sato 
\cite{sato81} derived an achievable capacity region (inner bound) of Gaussian 
interference channel as intersection of two multiple access gaussian capacity 
regions embedded in the interference channel. The achievable region is the 
subset of the rate pair  of the rectangular region of the very strong 
interference (1
\ref{VSI_RR}) and the region 
\beq
R_1+R_2 \le \log_2\left(\min\left\{1+P_1+\ga P_2,1+P_2+\gb P_1\right\} \right)
\eeq
A recent progress for the case of Gaussian interference is described by Sason  \cite{sason2004}.
Sason derived an achievable rate region based on a modified time- (or frequency-) division 
multiplexing approach which was originated by Sato for the degraded Gaussian IC. The achievable 
rate region includes the rate region which is achieved by time/frequency division multiplexing 
(TDM/ FDM), and it also includes the rate region which is obtained by time sharing between the 
two rate pairs where one of the transmitters sends its data reliably at the maximal possible 
rate (i.e., the maximum rate it can achieve in the absence of interference), and the other 
transmitter decreases its data rate to the point where both receivers can reliably decode their messages. 

In this paper we limit ourselves to the frequency-division multiplexing (FDM) scheme 
where an assignment of disjoint portions of the  frequency band to the several transmitters is 
made. This technique is widely used in practice because simple filtering can be used at the 
receivers to eliminate interference. The results equivalently apply to time-division multiplexing (TDM). In both cases we use the non-naive version, where all power is used in the frequency/time slice allocated for a given user.
 
While information theoretical considerations allow all points in the rate region, we argue that
the interference channel is a conflict situation 
between the interfering links. Each link is considered a player in a 
general interference game.  As such it has been shown that non-cooperative 
solutions such as the iterative water-filling, which leads to good solutions 
for the multiple access channel (MAC) and the broadcast channel \cite{yu04} 
can be highly suboptimal in interference  channel scenarios \cite{laufer2005},
\cite{laufer2005a}. To solve this problem 
there are several possible approaches. One that has gained popularity in 
recent years is through the use of competitive strategies in repeated games 
 \cite{etkin2005}, \cite{clemens2005}. Our approach is significantly 
 different and is based on general bargaining theory originally developed by 
Nash.  We claim that while all points on the boundary of the interference 
channel are achievable from  the  strict informational point of view, most of 
them will never be achieved since  one of the players will refuse to use coding 
strategies leading to these points.
 The rates of interest are only rates that are higher than the rates 
 that each user can achieve, independently of the other user coding  strategy. 
Such a rate pair must form a Nash equilibrium \cite{nash51}. 
This implies that not all the rates achievable from pure information theoretic 
point of view are indeed achievable from game theoretic prespective. Hence we define the game theoretic 
rate region.
\bdf
Let $\cR$ be an achievable information theoretic rate region. The game theoretic rate region ${\cR}^G$ is given by  
 \beq
 \cR^G=\left \{ 
 (R_1,R_2)\in \cR: R^c_i \le R_i, \ \ i=1,2 \right\}
 \eeq
 where $R^c_i$ is the rate achievable by user $i$ in a non-cooperative 
interference game.
\edf
 To see what are the pair rates that can be achieved by negotiation of the two 
users we resort 
 to a well known solution termed the Nash bargaining 
 solution. In his seminal papers \cite{nash50}, \cite{nash53}, Nash proposed 4 axioms that a 
 solution to a bargaining problem should satisfy. He then proves that there 
 exists a  unique solution satisfying these axioms. We will analyze the 
 application of Nash bargaining solution (NBS) to the interference game, and 
 show that there exists a unique point on the boundary of the capacity region 
 which is the solution to the bargaining problem as posed by Nash. 

 The fact that the Nash  solution can be computed independently by users, using 
 only channel state information, provides a good method for managing 
 multi-user ad-hoc networks operating in an unregulated environment. 

 Due to space limitations the paper considers only the Gaussian interference 
 channel and FDM strategy suitable for medium interference \cite{costa85}. 
 However extensions to other achievable rate regions of the interference channels will appear in  a subsequent paper.
 
Application of Nash bargaining to  OFDMA has been proposed by 
\cite{han2005}. However in that paper the solution was used only as a measure 
of fairness.  Therefore $R^c_i$ was not taken as the Nash equilibrium for the 
competitive game, but an arbitrary $R^{\min}_i$. This can result in 
non-feasible problem, and the proposed algorithm might be unstable. Furthermore
 our approach can be extended to other coding strategies such as in 
\cite{sason2004}.

\section{Nash equilibrium vs. Nash Bargaining solution}
In this section we describe to solution concepts for 2 players games. The 
first notion is that of Nash equilibrium. The second is the Nash bargaining 
solution (NBS). 
In order to simplify the notation we specifically concentrate on the Gaussian 
interference game.

 \subsection{The Gaussian interference game}
 \label{sec:GI_game} In this section we define the Gaussian
 interference game, and provide some simplifications for dealing with
 discrete frequencies. For a general background on non-cooperative
 games we refer the reader to \cite{owen} and \cite{basar82}.
 The Gaussian interference game was defined in
 \cite{yu2002}. In this paper we use the discrete approximation
 game. Let $f_0 < \cdots <f_K$ be an increasing sequence of
 frequencies. Let $I_k$ be the closed interval be given by
 $I_k=[f_{k-1},f_k]$. We now define the approximate Gaussian
 interference game denoted by $GI_{\{I_1, \ldots, I_K\}}$.

 Let the players $1,\ldots,N$ operate over separate channels. Assume
 that the $N$ channels have crosstalk coupling functions $h_{ij}(k)$.
 Assume that user $i$'th is allowed to transmit a total power of
 $P_i$. Each player can transmit a power vector $\vp_i=\left(
 p_i(1),\ldots,p_i(K) \right)  \in [0,P_i]^K$ such that $p_i(k)$ is
 the power transmitted in the interval $I_k$. Therefore we have $
 \sum_{k=1}^K p_i(k)=P_i$. The equality follows from the fact that in
 non-cooperative scenario all users will use the maximal power they
 can use. This implies that the set of power distributions for all
 users is a closed convex subset of the cube $\prod_{i=1}^N
 [0,P_i]^K$ given by: \beq \label{eq_strategies} \mB=\prod_{i=1}^N
 \mB_i \eeq where $\mB_i$ is the set of admissible power
 distributions for player $i$ is \beq \mB_i=[0,P_i]^K\cap
 \left\{\left(p(1),\ldots,p(K)\right): \sum_{k=1}^K p(k)=P_i \right\}
 \eeq Each player chooses a PSD $\vp_i=\left<p_i(k): 1\le k \le N
 \right > \in \mB_i$. Let the payoff for user $i$ be given by: \beq
 \label{eq_capacity}
 C^i\left(\vp_1,\ldots,\vp_N\right)= \\
 \sum_{k=1}^{K}\log_2\left(1+\frac{|h_i(k)|^2p_i(k)}{\sum
 |h_{ij}(k)|^2 p_j(k)+\vn(k)}\right)
 \eeq where $C^i$ is the capacity available to player $i$ given power
 distributions $\vp_1,\ldots,\vp_N$, channel responses $h_i(f)$,
 crosstalk coupling functions $h_{ij}(k)$ and $n_i(k)>0$ is external
 noise present at the $i$'th channel receiver at frequency $k$. In
 cases where  $n_i(k)=0$ capacities might become infinite using FDM
 strategies, however this is non-physical situation due to the
 receiver noise that is always present, even if small. Each $C^i$ is
 continuous on all variables.

 \begin{definition}
 The Gaussian Interference game $GI_{\{I_1,\ldots,I_k\}}=\left\{\mC,\mB\right\}$ is the N
 players non-cooperative game with payoff vector
 $\mC=\left(C^1,\ldots,C^N \right)$ where $C^i$ are defined in
 (\ref{eq_capacity}) and $\mB$ is the strategy set defined by (\ref{eq_strategies}).
 \end{definition}
 The interference game is a special case of non-cooperative N-persons
 game. 
\subsection{Nash equilibrium in non-cooperative games}
An important notion in game theory is that of a Nash
 equilibrium. 
 \bdf
 An $N$-tuple of strategies $\left<\vp_1,\ldots,\vp_N\right>$ for
 players $1,\ldots,N$ respectively
 is called a Nash equilibrium iff for all $n$ and for all $\vp$ ($\vp$ a
 strategy for player $n$)
 \[
 C^n\left(\vp_1,...,\vp_{n-1},\vp,\vp_{n+1},\ldots,\vp_N \right)<
 C^n\left(\vp_1,...,\vp_{N} \right)
 \]
 i.e., given that all other players $i \neq n$ use strategies $\vp_i$, player
 $n$ best response is $\vp_n$.
 \edf
 The proof of existence of Nash equilibrium in the general interference
 game follows from an easy adaptation of the proof of the this result
 for convex games \cite{nikaido55}.  
 A much harder problem is the uniqueness of Nash equilibrium points in
 the water-filling game. This is very important to the stability of the
 waterfilling strategies. A first result in this direction has been
 given in \cite{chung2002}. A more general analysis of the convergence
 (although it still does not cover the case of arbitrary channels) has
 been given in  \cite{luo2005}.

 \subsection{Nash bargaining solution for the interference game}
Nash equilibria are inevitable whenever non-cooperative zero sum
 game is played. However they can lead to substantial loss to all players, 
compared to a  cooperative strategy in the non-zero sum case, where players 
can cooperate. Such a situation is called the prisoner's dilemma.
The main issue in this case is how to achieve the cooperation in a stable 
manner and what rates can be achieved through cooperation.

In this section we present the Nash bargaining solution \cite{nash50}, 
\cite{nash53}. The underlying structure for a Nash bargaining in an $N$ 
players game is a set of outcomes of the bargaining process $S$ which is compact and convex. $S$ can be considered as a set of possible joint strategies or 
states, a designated 
disagreement outcome $d$ (which represents the agreement to disagree and solve 
the problem competitively) and a multiuser utility function 
$
U:S \cup \{d\} \rarrow {\bm R}^N.
$
The Nash bargaining is a function $F$ which assigns to each pair
$\left(S \cup \{d\},U \right)$ as above an element of $S \cup \{d\}$. 
Furthermore, the Nash solution is unique. In order to obtain the solution, Nash
 assumed four axioms:
\bds
\item[]{\em Linearity}. This means that if we perform  the same 
linear transformation on the utilities of all players than the solution is 
transformed accordingly.
\item[]{\em Independence of irrelevant alternatives}. This axiom states that 
if the bargaining solution of a large game $T \cup \{d\}$ is obtained in a 
small 
set $S$. Then the bargaining solution assigns the same solution to the smaller game, i.e., The irrelevant alternatives in $T \backslash S$ do not affect the outcome of the bargaining.
\item[]{\em Symmetry}. If two players are identical than renaming them will not 
change the outcome and both will get the same utility.
\item[]{\em Pareto optimality}. If $s$ is the outcome of the bargaining then no other state $t$ exists such that $U(s)<U(t)$ (coordinate wise).
\eds
A good discussion of these axioms can be found in \cite{owen}. 
Nash proved that there exists a unique solution to the bargaining problem 
satisfying these 4 axioms. The solution is obtained by maximizing
\beq
s= \arg \max_{s\in S \cup \{d\}} \prod_{n=1}^N \left(U_i(s)-U_i(d) \right)
\eeq
Typically one assumes that there exist at least one feasible 
 $s \in S$ such that $U(d)<U(s)$ coordinatewise, but otherwise we can assume that the bargaining solution is $d$.
In our case the utility for user $i$ is given by the rate $R_i$, 
and $U_i(d)$ is the competitive  Nash equilibrium, obtained  by 
iterative waterfilling for general ISI channels.

 \section{Existence and uniqueness of Nash bargaining solution for the two 
 players interference game}
 In this section we outline the proof that a Nash bargaining solution always 
 exists for utility function given by capacity for any achievable rate region 
 for the 2x2 interference channel. 

\begin{figure}
\begin{center}
\mbox{\psfig{figure=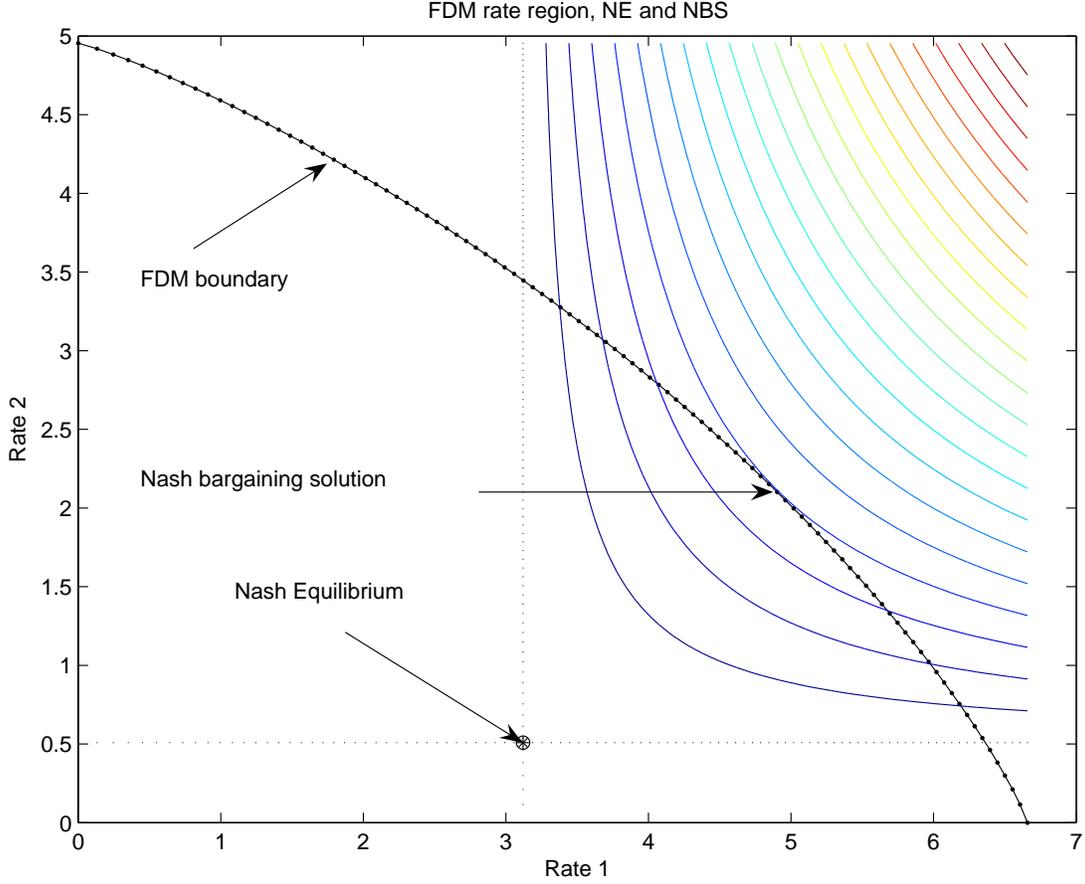,width=0.8\textwidth}}
\end{center}
    \caption{FDM rate region (thick line), Nash equilibrium $*$, Nash bargaining solution and the contours of $F(\gr)$. $\SNR_1=20$ dB, $\SNR_2=15$ dB, 
and $\ga=0.4, \gb=0.7$ }
\label{rate_region}
\end{figure}

An achievable rate region can always be 
defined by  the following equations:
 \beq
 \label{RR_constraint}
 \bea{l}
 0 \le R_1 \le R^1_{\max} \\
 0 \le R_2 \le f(R_1)
 \ena
 \eeq
 where $f(R)$ is a monotonically decreasing concave function of $R_1$. 
The monotonicity is obvious and the concavity follows from a standard time 
sharing argument.
 \bthm
\label{uniqueness}
Assume that we are given an achievable rate region for the interference channel
described by (\ref{RR_constraint}).
 Let $U_i(R)=R$ be the utility of the $i$'th user, and let $R^c_i$ be the 
 achievable rate at the competitive Nash equilibrium point for the $i$'th 
channel. Then there is a unique point $\left(R^{NBS}_1,R^{NBS}_2 \right)$
 that is the Nash bargaining  solution using the encoding strategies of the 
given achievable rate region for  the interference channel. This point is 
Pareto optimal and therefore on 
the boundary of the rate region, i.e., $R_2=f(R_1)$.
\ethm
Note that by concavity $f$ is strictly decreasing, except on an initial 
segment of rates for player I.
Due to space limitations, full proof of this theorem will be provided in an 
extended version of this paper. However the following example provides the 
intuition 
underlying   theorem \ref{uniqueness}, the relation between the competitive 
solution, the NBS and  
 the game theoretic rate region $\cR^G$ we have chosen $\SNR_1=20$ dB, 
$\SNR_2=15$ dB, 
and $\ga=0.4, \gb=0.7$.
Figure \ref{rate_region} presents the FDM rate region, the Nash equilibrium point denoted by 
$\*$, and a contour plot of $F(\gr)$. It can be seen that the convexity of $F(\gr)$ together 
with the concavity of the function defining the upper oundary of the rate region implies that at
there is a unique contour tangent to the rate region. The tangent point is the Nash bargaining 
solution. We can see that the NBS achieves rates that are 1.6 and 4 times higher than the 
rates of the competitive Nash equilibrium rates for player I and player II respectively. 
The game theoretic rate region is the intersection of
the information theoretic rate region with the quadrant above the dotted lines.
Finally we comment that this theorem can be generalized to $N$ players with higher notational complexity.
 \section{Bargaining for the two players interference game}
 In this section we analyze the two players interference game, with TDM/FDM 
 strategies. We provide conditions under which the bargaining solution provides 
 improvement over the competitive solution. This extends the work of 
 \cite{laufer2005} where it is characterized when does FDM solution outperforms
  the competitive IWF solution for symmetric 2x2 interference game. We assume 
 that the utility of player $i$ is given by $U_i=R_i$, the achievable rate. 
 To that end lets consider the general 2x2 interference channel (in 
 non-standard form). The received signal vector $\vx$ is given by
 \beq
 \vx=\mH \vs + \vn
 \eeq
 where $\vx=[x_1,x_2]^T$ is the received signal,
 \beq
 \mH=\left[
 \bea{cc}
 h_{11} & h_{12} \\
 h_{21} & h_{22}
 \ena
 \right]
 \eeq
 is the channel matrix and $\vs=[s_1,s_2]^T$ is the vector of transmited signals.
 Note that in our cases both transmission and reception are performed 
 independently, and the vector formulation is for notational simplicity only.
 Similarly to the analysis of \cite{laufer2005} the competitive strategies in 
 the interference game are given by flat power allocation resulting in rates:
 \beq
 \bea{l}
 R^c_{1}=\frac{W}{2}
 \log_2 \left(1+\frac{|h_{11}|^2 P_{1}}{WN_0/2+|h_{12}|^2 P_{2}}\right) \\
 R^c_2=\frac{W}{2}\log_2 \left(1+\frac{|h_{22}|^2 P_{2}}{WN_0/2+|h_{21}|^2P_{1}} \right)
 \ena
 \eeq
 Dividing by the noise power $WN_0/2$ we obtain
 \beq
\label{competitive}
 \bea{l}
 R^c_1=\frac{W}{2}\log_2 \left(1+\frac{\SNR_1}{1+\ga\SNR_2} \right) \\
 R^c_2=\frac{W}{2}\log_2 \left(1+\frac{\SNR_2}{1+\gb\SNR_1} \right) 
 \ena
 \eeq
 where 
 \[
 \bea{lclcl}
 \SNR_i=\frac{|h_{ii}|^2 P_i}{WN_0/2},&\ \ &
 \ga=\frac{|h_{12}|^2}{|h_{22}|^2},& \ \  &
 \gb=\frac{|h_{21}|^2}{|h_{11}|^2}
 \ena
 \]
This is equivalent ot the standard channel (\ref{standard_IC}), with $P_i=\SNR_i$.
 Since the rates $R^c_i$ are achieved by competitive strategy, player $i$ would not cooperate unless he will obtain a rate higher than $R^c_i$. Therefore the game theoretic rate region is defined by pair rates higher that $R_i^c$ of equation (\ref{competitive}).

Since we are interested in FDM cooperative strategies assume that player I uses a fraction of 
 $0 \le \gr \le 1$ of the band and user II uses a fraction $1-\gr$. The rates obtained by the two users are given by
 \beq
 \label{def_Rrho}
 \bea{l}
 R_1(\gr)=\frac{\gr W}{2} \log_2 \left(1+\frac{\SNR_1}{\gr} \right) \\
 R_2(1-\gr)=\frac{(1-\gr) W}{2} \log_2 \left(1+\frac{\SNR_2}{1-\gr} \right)
 \ena
 \eeq 
 The two users will benefit from FDM type of cooperation as long as 
 \beq
 \bea{l}
 R^c_i \le R_i(\gr_i), \ \ \ i=1,2 \\
 \gr_1+\gr_2 \le 1
 \ena
 \eeq
 For each $0<x,y$ define $f(x,y)$ as the minimal $\gr$ that solves the equation
 \beq
 \label{eq:def_f}
 \left(1+\frac{x}{\gr} \right)^\gr=1+\frac{x}{1+y}
 \eeq
\bcl
$f(x,y)$ is a well defined function for $x,y \in {{\bm R}^+}$.
\ecl
 Proof: Let 
 \[
 g(x,y,\gr)=\left(1+\frac{x}{\gr} \right)^\gr-1-\frac{x}{1+y}
 \]
 For every $x,y$, $g(x,y,\gr)$ is a continuous and monotonic function in $\gr$. 
 Furthermore, for any $0 \le x,y$, $g(x,y,1)>0$, while 
 \[
 \lim_{\gr \rarrow 0} g(x,y,\gr)<0.
 \]
 so there is a unique solution to (\ref{eq:def_f}).
\bcl
 Assume now that
 \beq
\label{FDM_better}
 f(\SNR_1,\ga \SNR_2)+f(\SNR_2,\gb \SNR_1) \le 1.
 \eeq 
 Then an FDM Nash bargaining solution exists. The NBS is given by solving the problem
 \beq
 \max_\gr F(\gr)=\max_{\gr}
 \eeq
 where 
\beq
\label{def_F}
F(\gr)=\left(R_1(\gr)-R^c_1 \right) \left(R_2(1-\gr)-R^c_2 \right)
\eeq
and $R_i(\gr)$ are defined by (\ref{def_Rrho}).
 \ecl
Proof: Let $\gr_1=f(\SNR_1,\ga \SNR_2),\gr_2=f(\SNR_2,\gb \SNR_1)$.
By definition of $f$ player $i$ has the same rate as the competitive rate if he can use
a $\gr_i$, fraction of the bandwidth. Since  (\ref{FDM_better}) implies that 
$\gr_1+\gr_2 \le 1$ FDM is preferable to the competitive solution.

A special case can now be derived:
 \bcl
 Assume that $\SNR_1 \ge \frac{1}{2} \left( \ga^2 \gb^4\right)^{-1/3}$ and
 $\SNR_2 \ge \frac{1}{2} \left( \gb^2 \ga^4\right)^{-1/3}$. Then there is a Nash bargaining solution that is better than the competitive solution. 
 \ecl
 The proof of the claim follows directly  by substituting $\gr_1=\gr_2=1/2$.

We also provide without proof the asymptotic performance as $\SNR_i$ 
increases to infinity 
\bcl
For each $i$ and for any fixed $z$, $\lim_{SNR_i \rarrow \infty}f(\SNR_i,z)=0$.
\ecl
This implies that if one of the users has sufficiently high $\SNR$ than 
FDM strategy is preferable to competitive strategy. 
This fact will be evident from the simulations in the next section.
\section{Simulations}
\begin{figure}
  \begin{center}
    \mbox{\psfig{figure=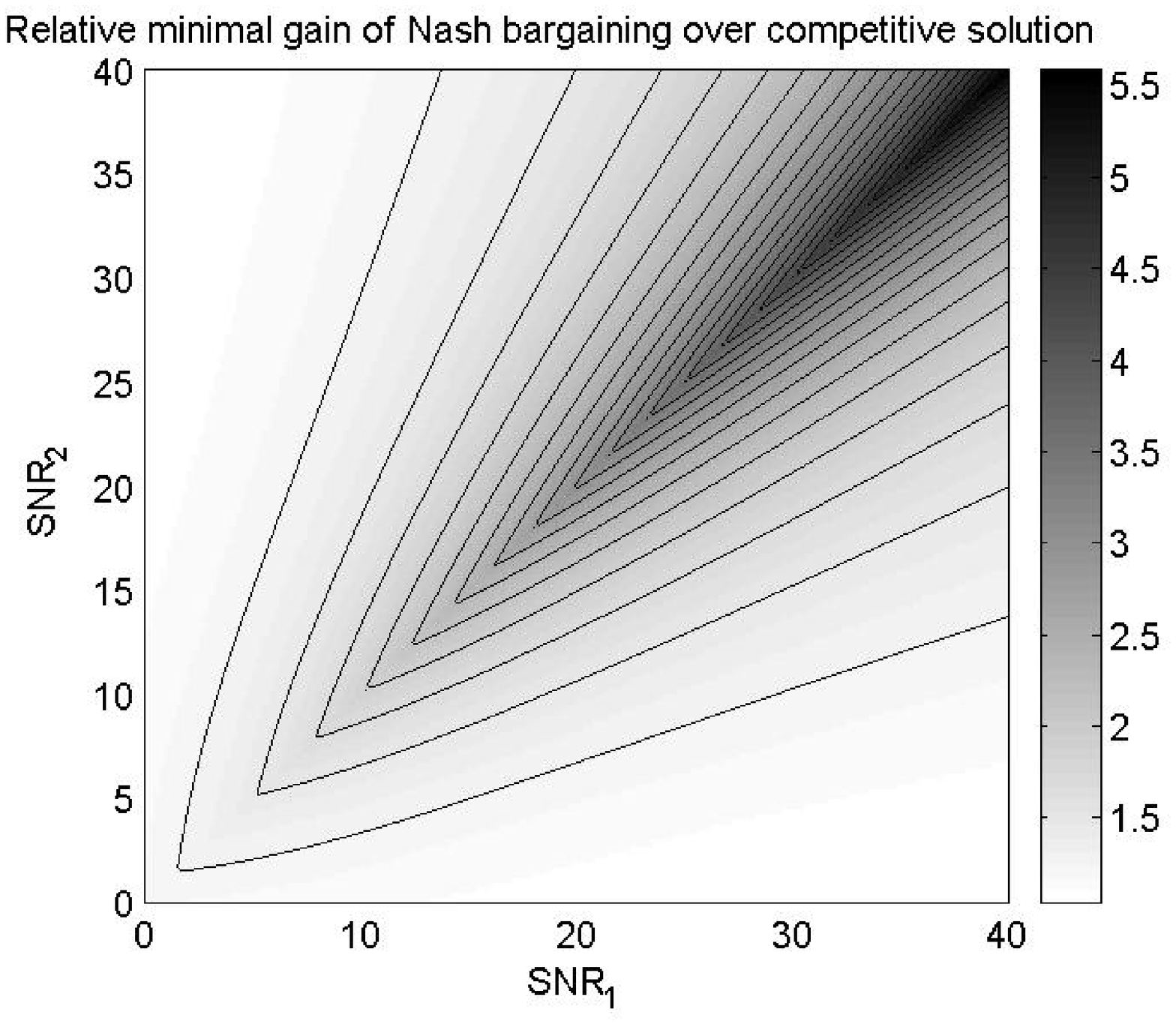,width=0.8\textwidth}}
 \end{center}
   \caption{Minimal relative improvement. $\ga=\gb=0.7$.}
   \label{snra7b7}
 \end{figure}

\begin{figure}
 \begin{center}
    \mbox{\psfig{figure=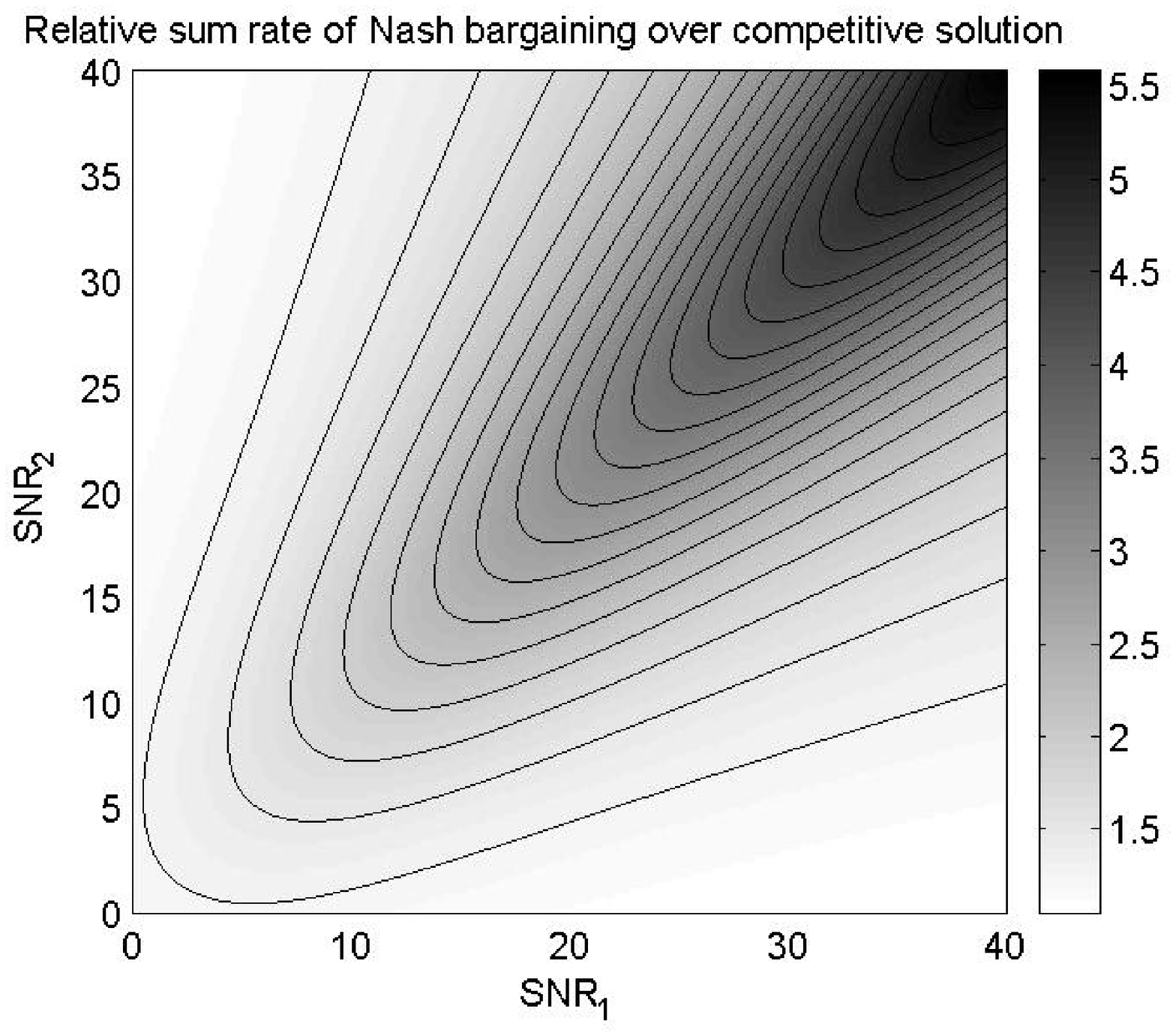,width=0.8\textwidth}}
 \end{center}
    \caption{sum rate relative improvement. $\ga=\gb=0.7$.}
 \label{sum_rate_SNRa7b7}
 \end{figure}

\begin{figure}
  \begin{center}
    \mbox{\psfig{figure=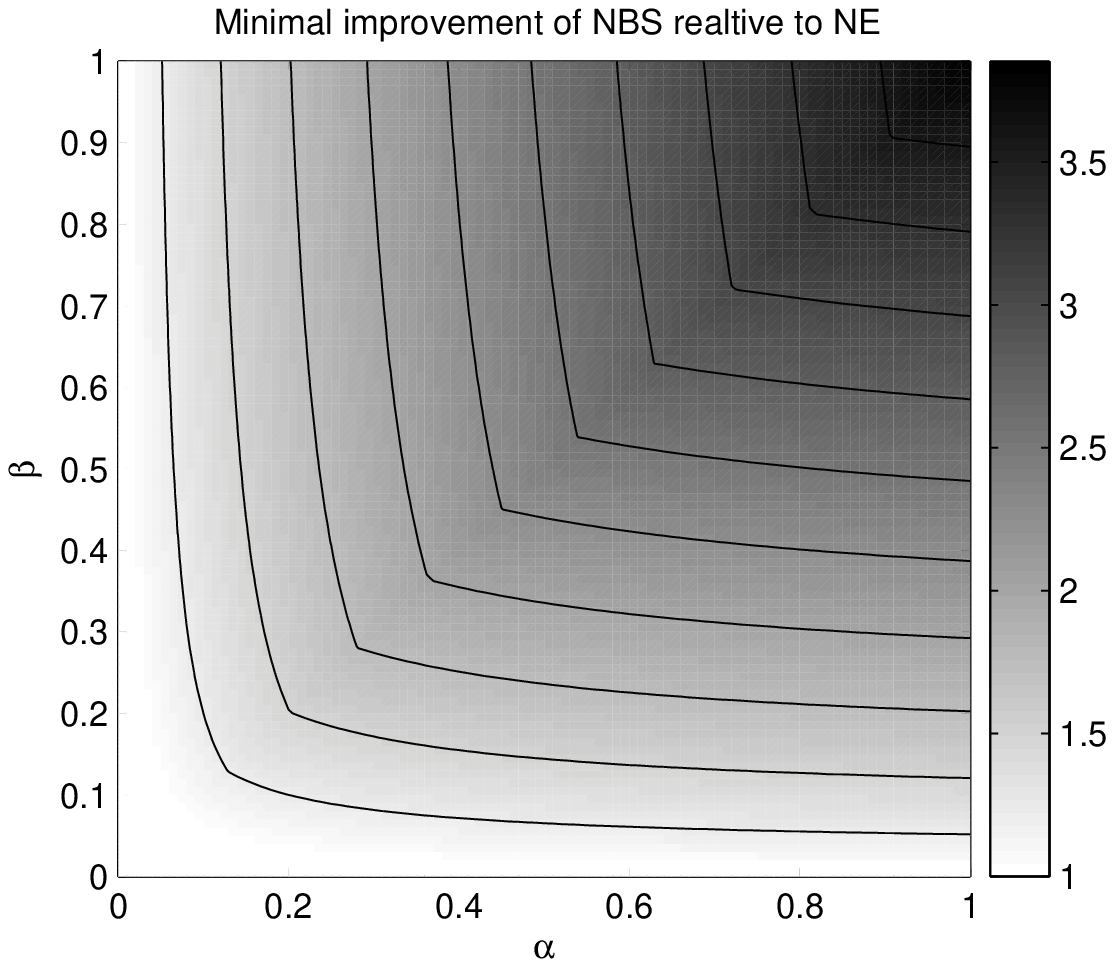,width=0.8\textwidth}}
 \end{center}
  \caption{Minimal relative improvement. SNR=20 dB.}
  \label{snr_minCH20}
 \end{figure}

\begin{figure}
  \begin{center}
    \mbox{\psfig{figure=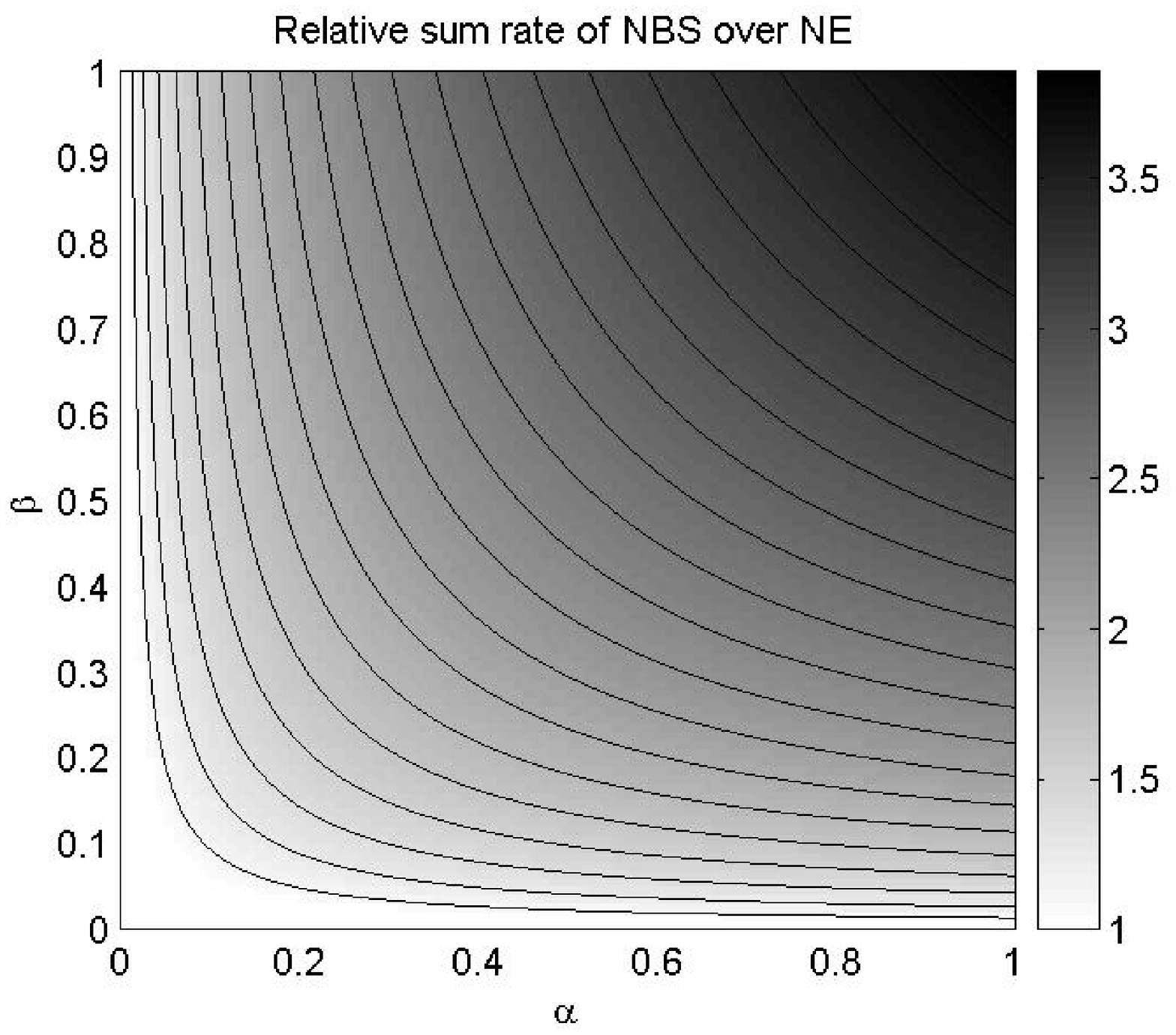,width=0.8\textwidth}} \end{center}
    \caption{Sum rate  relative improvement. SNR=20 dB.}
\label{sum_rate_snrCH20}
 \end{figure}

In this section we compare in simulations the Bargaining solution to the 
competitive solution for various situations with medium interference. 
We have  tested the gain of the Nash bargaining solution relative to the Nash 
 equilibrium competitive rate pair as a function of channel coefficients as 
 well as signal to noise ratio.  To that end we define the minimum relative 
improvement by:
 \beq
 \label{min_delta}
 \gD_{\min}=\min\left\{ 
 \frac{R^{NBS}_1}{R^c_1},\frac{R^{NBS}_2}{R^c_2}
   \right\}
 \eeq
 and the sum rate improvement by
 \beq
 \label{sum_delta}
 \gD_{sum}=\frac{R^{NBS}_1+R^{NBS}_2}{R^c_1+R^c_2}
 \eeq
 In the first set of experiments we have fixed $\ga,\gb$ and 
 varied $\SNR_1,\SNR_2$ from 0 to 40 dB in steps of 0.25dB. 
 Figure \ref{snra7b7} presents $\gD_{\min}$ for an 
 interference channel with $\ga=\gb=0.7$. We can see that for high 
 SNR we obtain significant improvement. Figure \ref{sum_rate_SNRa7b7} 
 presents the relative sum rate improvement $\gD_{sum}$ for the same channel. 
 We can see that the achieved rates are 5.5 times those of the competitive 
 solution. 
 We have now studied the effect of the interference coefficients on the Nash 
 Bargaining solution. We have set the signal to additive white Gaussian noise 
 ratio for both users to 20 dB, and varied $\ga$ and $\gb$ between $0$ and $1$. 
 Similarly to the previous case we present the minimal improvement 
$\gD_{\min}$ and the sum rate improvement $\gD_{sum}$. The results are shown 
in figures \ref{snr_minCH20},\ref{sum_rate_snrCH20}.

 \section{Conclusions}
In this paper we have defined the game theoretic rate region for the 
interference channel. The region is a subset of the rate region of the 
interference channel. We have shown that a specific point in the rate region
given by the Nash bargaining solution is better than other points in the context of bargaining theory. We have shown conditions for the existence of such a 
point in the case of the FDM rate region. Finally we have demonstrated through 
simulations the significant improvement of the cooperative solution over the 
competitive Nash  equilibrium.

 \bibliographystyle{ieeetr}
\newcommand{\noopsort}[1]{} \newcommand{\printfirst}[2]{#1}
  \newcommand{\singleletter}[1]{#1} \newcommand{\switchargs}[2]{#2#1}

 \end{document}